\documentclass[letter,preprint]{jpsj3}
\usepackage{txfonts}

\title
{
Comparative Study of First and Second Switches 
with Nearly Identical Switching Current 
in Bi$_2$Sr$_2$Ca$_{0.85}$Y$_{0.15}$Cu$_2$O$_y$ Intrinsic Josephson Junctions 
}

\author{
Ayami Yamaguchi, Haruka Ohnuma, Yuji Watabe, Shumpei Umegai, Kazutaka Hosaka, and Haruhisa Kitano\thanks{e-mail: hkitano@phys.aoyama.ac.jp}
}
\inst{Department of Physics and Mathematics, Aoyama Gakuin University \\
5-10-1 Fuchinobe, Chuo-ku, Sagamihara, Kanagawa 252-5258, Japan
} 

\abst{
We present a study of the first and second switches (SWs) 
with nearly identical switching current $I_{\rm SW}$ 
in Bi$_2$Sr$_2$Ca$_{0.85}$Y$_{0.15}$Cu$_2$O$_y$ intrinsic Josephson junctions (IJJs). 
In contrast to early studies where a large difference between $I_{\rm SW}^{\rm 1st}$ and $I_{\rm SW}^{\rm 2nd}$ was observed, a crossover temperature $T_{\rm cr}$ to the quantum phase escapes for the 2nd SW is strongly reduced $(\sim$2.8~K) and agrees with that for the 1st SW. This confirms that the anomalous enhancement of $T_{\rm cr}^{\rm 2nd}$, commonly observed in the IJJs with $I_{\rm SW}^{\rm 1st} < I_{\rm SW}^{\rm 2nd}$, originates from the phase running state after the 1st SW. 
Detailed analysis of the microwave-induced escapes strongly suggests that nonlinear bifurcated phenomena possibly occurs in IJJs. 
}

\begin{document}
\maketitle

Macroscopic quantum tunneling (MQT) phenomena discovered in the intrinsic Josephson junctions (IJJs) of Bi$_2$Sr$_2$CaCu$_2$O$_y$ (Bi2212) superconductors \cite{Inomata} have attracted much interest for more than a decade because of the unusual properties on the MQT rate. \cite{Jin,Severev,Fistul2,Koyama} 
The most striking feature is that a crossover temperature, $T_{\rm cr}$, to the MQT state in the phase switch from the first to the second resistive branch (2nd SW) is largely enhanced compared to that from the zero voltage to the first resistive branch (1st SW) in the multi-branched current-voltage ($I$-$V$) characteristics of IJJs  \cite{Kashiwaya,Ota,Warburton,Nomura,Kakehi,Kitano,Takahashi,Nomura2,Kitano2,Kitano3}. 
An early study \cite{Kashiwaya} suggested that such an enhancement was due to the self-heating effects in the first resistive state, based on the consideration that a series array of uniform Josephson junctions should show the same value of $T_{\rm cr}$ as long as the current flowing in each junction agrees with the externally-applied bias current. 

However, the anomalous increase of $T_{\rm cr}^{\rm 2nd}$ was found to be independent of the device structure of IJJs with different heat-transfer environment \cite{Ota} and the order of phase switches, \cite{Kakehi} throwing doubt on the self-heating effects after the 1st SW. 
Detailed analyses of the switching rate for the 1st to the 4th SWs suggested that the phase-running state in the resistive branches enhances the phase retrapping effects in the higher order SWs rather than the self-heating effects. \cite{Kitano}
In addition, the microwave-induced resonant peak structure was observed in the switching current distribution below $T_{\rm cr}^{\rm 2nd}$, suggesting the occurrence of the quantum phase escapes and the formation of energy level quantization (ELQ) in a potential well. \cite{Takahashi,Kitano2}
Thus, many experimental results \cite{Ota,Nomura,Kakehi,Kitano,Takahashi,Kitano2,Nomura2,Kitano3} support that the MQT state is realized even in the 2nd SWs of IJJs, although the value of $T_{\rm cr}$ (typically $\sim$ 10~K for the optimal doped Bi2212 IJJs) is much higher than the value predicted in the conventional theory. \cite{Caldeira}

As a qualitative explanation for this anomaly, we proposed that the ac Josephson current occurring after the 1st SW possibly passes through other junctions near the phase-switched junction, inducing the rapid oscillation in a tilted-washboard potential and effectively reducing the barrier height $\Delta U(I)$ for the 2nd SW. \cite{Kitano}
Similar effects are expected for the quantum phase escapes under the strong microwave irradiation. 
Fistul {\it et al} reported that the strong microwaves effectively reduced $\Delta U(I)$ and increased the MQT rate, by using the quantum-mechanical analyses of the measured results for the Nb/AlO$_x$/Nb Josephson junctions.   \cite{Fistul}
These effects are also confirmed for Bi2212-IJJs under the strong microwaves. \cite{Takahashi,Kitano2,Kitano3,HFYu}
Thus, the alternating current generating after the 1st SW is also expected to increase the MQT rate due to the effective reduction of $\Delta U(I)$.
This scenario qualitatively explains the anomalous enhancement of $T_{\rm cr}$ for the 2nd and higher order SWs. 
 
An important fact leading to the above proposal is that the switching current $I_{\rm SW}$ for the 1st SW often becomes smaller than that for the 2nd SW ($I_{\rm SW}^{\rm 1st} < I_{\rm SW}^{\rm 2nd}$), because of the proximity effect of the normal-metal electrode in the mesa-type IJJs, \cite{HFYu,Nomura,Nomura2,SXLi} or the amorphous damage due to focused ion beam (FIB) microfabrication in the bridge-type IJJs \cite{Kakehi,Takahashi,Kitano3,Kakizaki}. 
The reduced switching current brings the phase-running state after the 1st SW, generating a temporal oscillation in a tilted washboard potential to describe the 2nd SW. 
This suggests that a study on IJJs with almost identical $I_{\rm SW}^{\rm 1st}$ and $I_{\rm SW}^{\rm 2nd}$ will provide novel information. 
A few studies reported the phase switching dynamics for such IJJs \cite{Warburton,Nomura3}, while there was no discussion about the above scenario. 

In this paper, we present a comparative study of the 1st and 2nd SWs in stacked IJJs made of 
Bi$_2$Sr$_2$Ca$_{0.85}$Y$_{0.15}$Cu$_2$O$_y$, where $I_{\rm SW}^{\rm 1st}$ and $I_{\rm SW}^{\rm 2nd}$ are nearly identical. 
We confirm that an effective escape temperature $T_{\rm esc}$ for the 2nd SW shows almost the same temperature dependence as that for the 1st SW, indicating that $T_{\rm cr}$ for the 2nd SW is not enhanced but close to that for the 1st SW ($\sim$2.8~K). 
This result sharply contrasts to the previous results for IJJs with $I_{\rm SW}^{\rm 1st} \ll I_{\rm SW}^{\rm 2nd}$, strongly supporting that the anomalous enhancement of $T_{\rm cr}^{\rm 2nd}$ is attributed to the influence of the phase-running state after the 1st SW. 
We also observe that the microwave-induced resonant escapes for both SWs manifest almost the same behaviors up to $\sim$15~K. 
The detailed analyses indicate that the decrease of the Josephson plasma frequency $\omega_p$ with increasing the microwave power is stronger than that of $\Delta U$, in contrast to the previous studies \cite{Takahashi,Kitano2}. 
Furthermore, the energy level numbers within the potential well for both SWs was found to remain more than 20 under the strong microwaves.   
These results suggest that the microwave irradiation is not used for a multi-photon absorption process from the ground state to the first excited state, but induces the strong oscillation such as a nonlinear bifurcation in the potential well.   
We argue that this unusual behavior gives an interesting key to resolve the anomalous enhancement of $T_{\rm cr}$ for the 2nd and higher order SWs.  

Y-doped  Bi2212 crystals were grown by the floating zone method and annealed in the same condition as the optimal doped Bi-2212. \cite{Kakehi,Kitano3}
A partial substitution of Y$^{3+}$ for Ca$^{2+}$ (nominally, 15$\%$) decreases the carrier concentration to the underdoped region with the superconducting transition at $\sim$65~K. 
A stack of IJJs was fabricated  in a narrow bridge by using FIB milling techniques. \cite{Kakehi} 
The lateral sizes and the thickness of the measured IJJs (ayf46) are 1.31$\times$1.39~$\mu$m$^2$ and 0.14 $\mu$m, respectively. 
The switching current distribution $P(I)$ was measured down to 0.8~K by applying a bias current with a constant rate $dI/dt=3$~mA/s under 5,000 or 10,000 switching events. 
Other details of experimental setup are described in our previous papers \cite{Ota,Takahashi}. 
The measured data of $P(I)$ and the corresponding switching rate $\Gamma (I)$ as a function of bias current were analyzed by a single-junction model including both the thermally activated (TA) phase escapes and the multiple phase retrapping (PR) processes, based on the so-called resistively and capacitively shunted junction (RCSJ) model. \cite{Kitano,Kitano2}
Other details including experimental methods to distinguish between each junction are described in a supplementary material. \cite {Memo1}   

Figure 1 shows $P(I)$ for the 1st and 2nd SWs measured at several temperatures from 0.8~K to 20~K. 
Here, open circles and solid lines are the experimental data and the numerical fittings, respectively. 
We find that the temperature dependences of $P(I)$ for both SWs are very similar, 
showing negligible temperature dependence for both SWs below 1.2~K.
On the other hand, the width of $P(I)$ for the 2nd SW becomes narrower than that for the 1st SWs above 8~K. 
The slight differences between the 1st and 2nd SWs are more clearly observed in the bias current dependence of $\Gamma (I)$, as shown in Figs. 2(a) and 2(b). 
In these plots, we directly compare $\Gamma(I)$ for the 1st SW (open circles) with those for the 2nd SW (solid triangles). 
$\Gamma(I)$ for both SWs agree with each other below 4~K, indicating that the switching current distribution for the 1st SW almost overlaps that for the 2nd SW.
On the other hand, $\Gamma(I)$ for the 2nd SW above 6~K shows a negative curvature in logarithmic plots and is located at a higher current region than that for the 1st SW. 
These results suggest that the multiple PR processes are increased for the 2nd SW above 6~K. 
This behavior has also been observed in the higher order SW of IJJs with the reduced  $I_{\rm SW}^{\rm 1st}$ \cite{Kitano,Kitano3}. 

Figures 2(c) and 2(d) show the temperature dependences of $T_{\rm esc}$ and a fluctuation-free critical current $I_{c{\rm 0}}$, respectively, which were obtained by the numerical fitting of $P(I)$ and $\Gamma(I)$. 
In contrast to the previous results, \cite{Kashiwaya,Ota,Nomura,Kakehi,Takahashi,Nomura2,Kitano3} 
the plots of $T_{\rm esc}$ vs $T$ for the 2nd SW agree well with those for the 1st SW from 0.8~K to  20~K. The corresponding behavior of $T_{\rm esc}$ for both SWs is expected for an ideal straight array of Josephson junctions, where the same bias current flows through each junction.  
This provides a strong piece of evidence that most of the previous results were derived from the difference between $I_{\rm SW}^{\rm 1st}$and $I_{\rm SW}^{\rm 2nd}$, and supports that the existence of the moderately-long running state after the 1st SW is crucial for the understanding of anomalous behaviors of $T_{\rm esc}$ for the 2nd SW.  
On the viewpoint of the Josephson radiation, which is generated in the voltage state of IJJs and moderately damped by a poor screening in the thin superconducting layers, our result suggests that the frequency chirp of the Josephson radiation is a key point, since a duration of the voltage state determines an extent of the chirp. 

The values of $I_{c0}$ for both SWs are 13.3~$\mu$A at 0.8~K, showing no temperature dependence up to 6~K. 
At temperatures above 8~K, $I_{c0}$ for the 2nd SW slightly becomes larger than that for the 1st SW, presumably due to the above-mentioned slight differences  between both SWs in $P(I)$ and $\Gamma(I)$.  
A critical current density $j_{c{\rm 0}}$ is estimated as 736~A/cm$^2$ for both SWs at 0.8~K, which is roughly 1/5 $\sim$ 1/4 of the obtained values for the optimally doped Bi2212-IJJs. 
Such a strong suppression of $j_{c{\rm 0}}$ by underdoping is also reported in the previous study of the interlayer tunneling spectroscopy with the mesa-type IJJs. \cite{Suzuki} 

It is interesting to note that the experimental value of $T_{\rm cr}$ for both SWs is about 2.8~K, which is much higher than the reported values for the 1st SW in Bi2212-IJJs, \cite{Inomata,Jin,Kashiwaya,Ota} while much lower than those for the 2nd SW.  \cite{Kakehi,Takahashi,Nomura2,Kitano3} 
The decrease of $T_{\rm cr}^{\rm 2nd}$ is consistent with our scenario that the phase escape after the 1st SW is strongly influenced by the phase-running state occurring in the switched junction, since a short period of the running state before the 2nd SW is not enough to affect the 2nd SW. 
On the other hand, the increase of $T_{\rm cr}^{\rm 1st}$ is quite unexpected, since $j_{c{\rm 0}}$ is strongly suppressed. 
This implies the importance of quantum effects, which may be derived from indistinguishable properties between neighboring junctions with identical switching current. \cite{Memo2}
Perhaps, this unusual enhancement of $T_{\rm cr}$ is related to the drastically-enhanced MQT rate observed in the uniformly switching Bi2212-IJJs. \cite{Jin,Severev,Fistul2,Koyama}
We also find that $T_{\rm esc}$ from 2.8~K to 6~K deviates from a straight line representing that $T_{\rm esc}=T$. 
This suggests that $\Gamma(I)$ in the regime that $T_{\rm cr} < T < 2.1~T_{\rm cr}$ is enhanced by quantum correction effects, which have  been considered in the conventional MQT theory for dissipative systems \cite{Cleland,Grabert}.  
These results also suggest that the MQT phenomena occurring in the IJJ system are still unresolved. The upper limit of $T_{\rm cr}$ attained in IJJs can be more enhanced.  

The formation of ELQ in the MQT regime is often investigated by the double-peaked distribution of the switching current under microwave irradiation, since the microwave absorption due to single- or multi-photon process stimulates the phase escape from the excited energy levels. \cite{Jin,Takahashi,Martinis} 
Figure 3 shows the density plots of the switching current for both SWs versus the applied microwave power $P_{\rm MW}$ at 21.4~GHz and at several temperatures up to 15~K. 
The left and right panels represent the 1st SW and the 2nd SW, respectively. 
First of all, we have an interest in the common results to both SWs. 
As the microwave power is increased at $T=$ 1.7~K below $T_{\rm cr}$, the distribution of $P(I)$ becomes double-peaked near $P_{\rm MW}=-$15~dBm ($-$13~dBm) for the 1st (2nd) SW. 
The applied microwave frequency corresponds to the four-photon process ($n=$ 4) for both SWs, assuming that the resonant peak is derived by the phase escape from the first excited state of ELQ. \cite{Kopietz} 
With increasing temperatures, the primary peak positioned in the lower power region moves downward and approaches the microwave-induced secondary peak. At 15~K, the double peak structure vanishes and the primary peak is  smoothly connected to the secondary peak.  
Note that the double peak in $P(I)$ is observed for both SWs even at temperatures above $T_{\rm cr}$. A similar behavior has also been reported in the previous studies. \cite{Takahashi,Takahashi_spl} 

Next, we focus on the detailed difference between both SWs.  
As shown in Figs.~3(a) and 3(e), we find that the microwave power required for the resonant phase escape in the 2nd SW is larger than that in the 1st SW. 
Here, we note that a small discontinuity of the switching current for the 2nd SW, observed at around $-$14.5~dBm, is not due to  the resonant escape. Rather it incidentally occurs by a large reduction of the switching current for the 1st SW. 
Thus, we deduce that a large reduction of the switching current observed at $P_{\rm MW}\sim-$13~dBm is a sign of the resonant phase escape for the 2nd SW.   
This irradiation power is stronger than the power ($\sim-$15~dBm) for the 1st SW.  
Such a difference between both SWs can be successfully explained by the large reduction of the switching current occurring in the 1st SW, which keeps the phase-running state after the 1st SW going for a longer time. 
This generates the ac Josephson current giving a stronger oscillation in the washboard potential to describe the 2nd SW.
Thus, the stronger irradiation power is needed to induce the resonant phenomena in the 2nd SW.   

We also performed the numerical fittings of $P(I)$ and $\Gamma(I)$ for both SWs under the microwave irradiation, by using the single-junction model used in the analyses without irradiation. 
In particular, we focus on the changes of $\Delta U(I)$ and the Josephson plasma frequency, $\omega_p(I)$ ($=\sqrt{2eI_{\rm c0}/\hbar C}[1-(I/I_{\rm c0})^2]^{1/4}$), where $C$ ($\sim$90~fF) is a capacitance for each junction in IJJs, as a function of microwave irradiation power. 
Figures 4(a) and 4(b) show the plots of $T_{\rm esc}$ and $I_{\rm c0}$ versus $P_{\rm MW}$ for both SWs, which were obtained from the measured results at $T=$~1.7~K and 3.8~K. We confirm that $T_{\rm esc}$ and $I_{\rm c0}$ are almost identical between both SWs at each temperature, showing a good agreement with the results obtained without the microwaves. On the other hand, the microwave power dependence of $T_{\rm esc}$ and $I_{\rm c0}$ is contrasted with each other. 
$T_{\rm esc}$ is almost independent of $P_{\rm MW}$ while $I_{\rm c0}$ is decreased with increasing $P_{\rm MW}$ at both temperatures. 
Such a reduction of $I_{\rm c0}$ with increasing $P_{\rm MW}$ was also reported in the previous study on the microwave-induced phase switches in the 2nd SW of Bi2212-IJJs, where the switching current for the 1st SW is smaller than that for the 2nd SW. \cite{Takahashi,Kitano2}

Figures 4(c) and 4(d) show the similar plots of $\Delta U(P_{\rm MW})$ and $\omega_p(P_{\rm MW})$, which are normalized by the values at $P_{\rm MW}=-$30~dBm to compare the change rates with each other.   
Note that the values of $T_{\rm esc}$ and $I_{\rm c0}$ at $P_{\rm MW}=-$30~dBm are almost the same as those obtained without irradiation. 
Both of $\Delta U$ and $\omega_p$ are slightly decreased with increasing $P_{\rm MW}$, and there is no difference between both SWs. 
On the other hand, the decrease of $\omega _p$ (roughly, $\sim$10 \%) is much larger than that of $\Delta U$ ($\sim$1-2 \%), showing a sharp contrast to the model proposed by Fistul {\it et al} \cite{Fistul}.
Furthermore, such an insensitivity of $\Delta U$ to $P_{\rm MW}$ is quite different from our previous results for the 2nd SW occurring at $I_{\rm SW}$ far away from the 1st SW. \cite{Takahashi, Kitano2} 
In the latter case, $\Delta U$ continues to decline until the resonant escapes are observed.  
On the other hand, $\omega _p$ initially shows a rapid decrease by half and changes into a moderate one, which is weaker than the decrease of $\Delta U$ near the resonant escapes. 
These results indicate that the microwave-induced resonant escapes observed in this work have a different origin from those observed in the previous studies. 

In connection with this discrepancy, the energy level numbers within the potential wells, calculated by using the previous approach, \cite{Kopietz} are too many to escape from the first excited state through the multi-photon absorption processes. 
The obtained values ($\gtrsim 20$ at $T=$1.7~K for both SWs) are roughly twice larger than the previous results. \cite{Kitano2,HFYu} 
This strongly suggests that the escaped phase particle is excited to a much-higher energy level, by using successive absorption of many photons or classical oscillation with extremely large amplitude. 
Although the former case is not reported yet as far as we know, the latter case has been reported as a nonlinear bifurcation phenomenon observed in the conventional Josephson junctions. \cite{HFYu2}
We note that the significant reduction of $I_{\rm SW}$ observed at the resonance (See also Fig.~3) is successfully explained by the contribution of the nonlinear bifurcated oscillation. 
If it actually occurs in the IJJs, the observed resonant escapes will not be always associated with the formation of ELQ.    
Nevertheless, the invariant behavior of $T_{\rm esc}$ with increasing $P_{\rm MW}$ seems to support the occurrence of MQT after the excitation to a high energy level by using the bifurcated oscillation. 

Our results also imply that a similar bifurcated oscillation possibly occurs in the 2nd and higher order SWs for the IJJs with $I_{\rm SW}^{\rm 2nd} \gg I_{\rm SW}^{\rm 1st}$, since the ac Josephson current generated in the phase running state plays a role similar to the microwave irradiation. 
This means that the nonlinear bifurcation possibly gives another answer to the anomalous enhancement of $T_{\rm cr}$ in the 2nd and higher order SWs. 
Note that this interpretation is not obtained by the Josephson radiation, where its frequency is chirped upward with increasing voltages. The up-chirp radiation does not seem to induce the nonlinear bifurcation suggested in this work, since it is usually observed by the down-chirp pulses in the conventional nonlinear systems. 
Thus, it is interesting to clarify which scenario between the nonlinear bifurcation or the chirped Josephson radiation explain the anomalous enhancement of $T_{\rm cr}$ more successfully. We will discuss about this issue in another paper in the near future. 

In summary, 
we experimentally studied the 1st and 2nd SWs in the Y-doped Bi2212-IJJs, where $I_{\rm SW}^{\rm 1st}$ and $I_{\rm SW}^{\rm 2nd}$ were nearly identical. 
We confirm that the behaviors of $T_{\rm esc}$ and $I_{\rm c0}$ observed for the 2nd SW are coincident with those for the 1st SW, which is contrast to the previous studies on Bi2212-IJJs with $I_{\rm SW}^{\rm 2nd}$ larger than $I_{\rm SW}^{\rm 1st}$.
From these contrasting results, we conclude that the anomalous enhancement of $T_{\rm cr}$ for the 2nd and higher order SWs is essentially derived from the phase-running state after the phase switches. 
As a possible mechanism for such an enhancement, the effective reduction of $\Delta U$ for the 2nd SW, \cite{Kitano, Takahashi} attributed to the ac Josephson current after the 1st SW, was proposed. 
This scenario based on the model proposed by Fistul {\it et al} \cite{Fistul} seemed to explain the observed results successfully. 
However, this work suggests that the microwave irradiation decreases $\omega_p$ rather than $\Delta U$ for both SWs, retaining many energy level numbers in the potential well.  
Thus, the microwave-induced resonance observed in this work is presumably attributed to the nonlinear bifurcated oscillation.  
It also implies that a similar bifurcated oscillation can be induced by the ac Josephson current after the 1st SW. 
The nonlinear bifurcation effects provide an important viewpoint to understand the complicated phase dynamics in IJJs, including the comprehension of the anomalous enhancement of $T_{\rm cr}$ for the 2nd and higher order SWs. 

We thank I. Kakeya for fruitful discussion. 
This work was partly supported by MEXT-Supported Program for the Strategic Research Foundation at Private Universities (2013-2017).
FIB microfabrication performed in this work was supported by Center for Instrumental Analysis, College of Science and Engineering, Aoyama Gakuin University.   


\newpage
\begin{figure}[t]
\includegraphics[width=70mm]{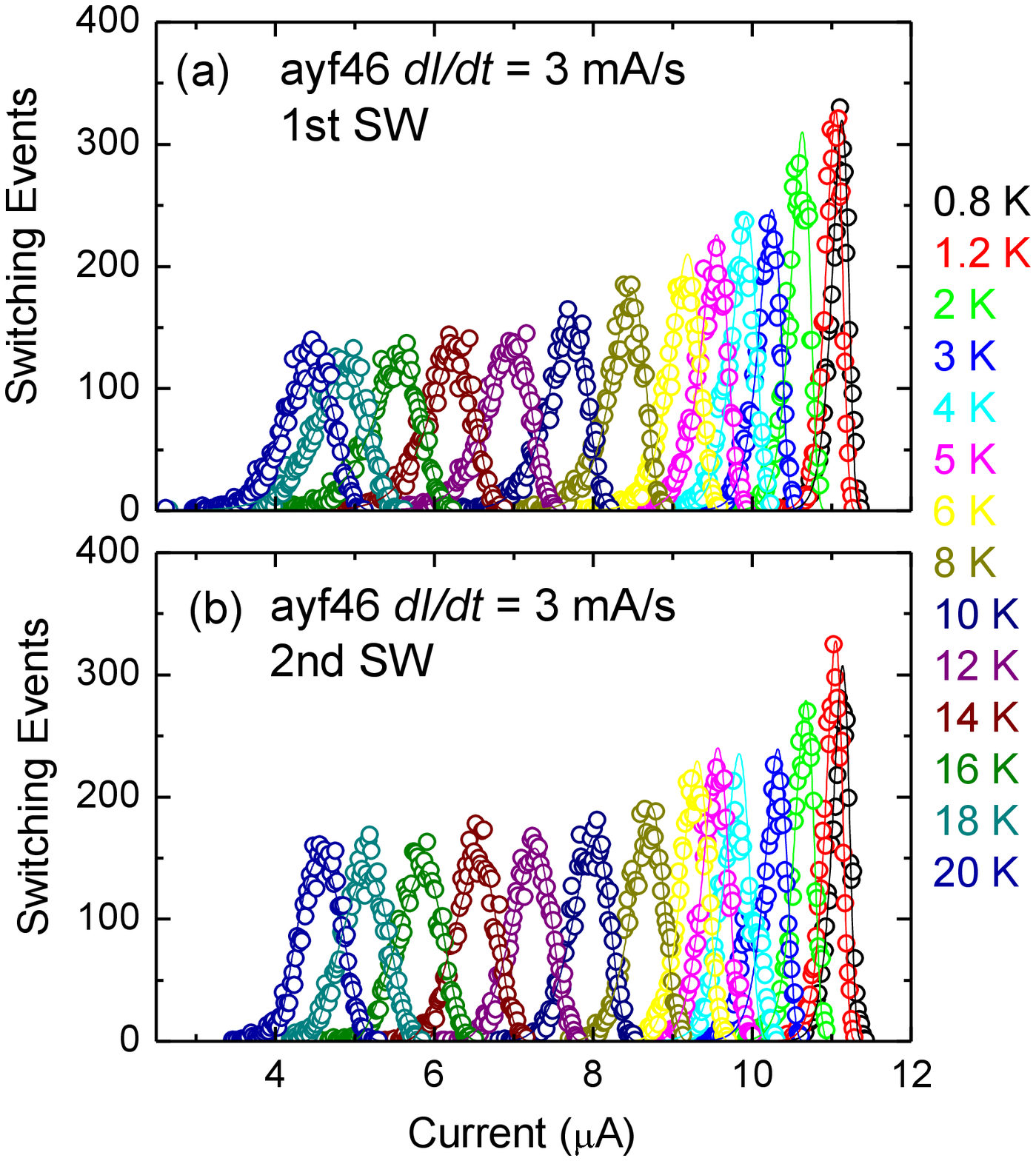}
\caption{
(Color online) (a) Switching current distribution for (a) the 1st SW and (b) the 2nd SW, measured at several temperatures from 0.8~K to 20~K. Open circles represent the experimental data, and solid lines represent the numerical fitting lines by TA+PR model. }
\label{f1}
\end{figure}

\begin{figure}[t]
\includegraphics[width=70mm]{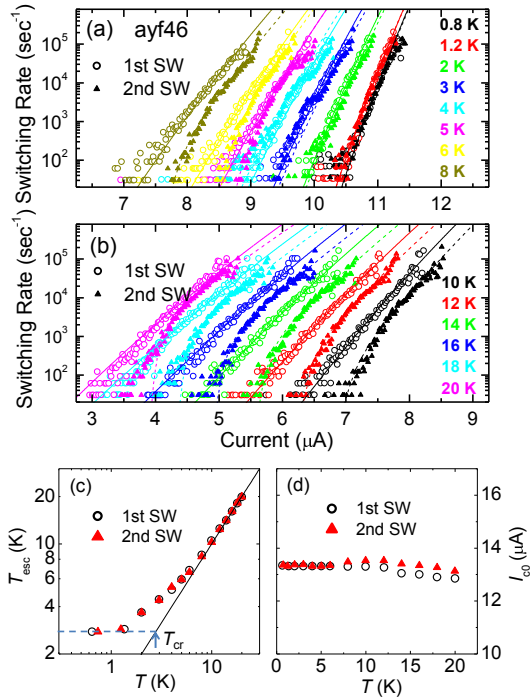}
\caption{
(Color online) Bias current dependence of the switching rate for the 1st and 2nd SWs at several temperatures: (a) 0.8~K to 8~K, (b) 10~K to 20~K, respectively, and the temperature dependence of (c) the effective escape temperature $T_{\rm esc}$ and (d) the fluctuation-free critical current $I_{\rm c0}$. Open circles and solid triangles represent the 1st SW and the 2nd SW, respectively. }
\label{f2}
\end{figure}

\begin{figure}[h]
\includegraphics[width=70mm]{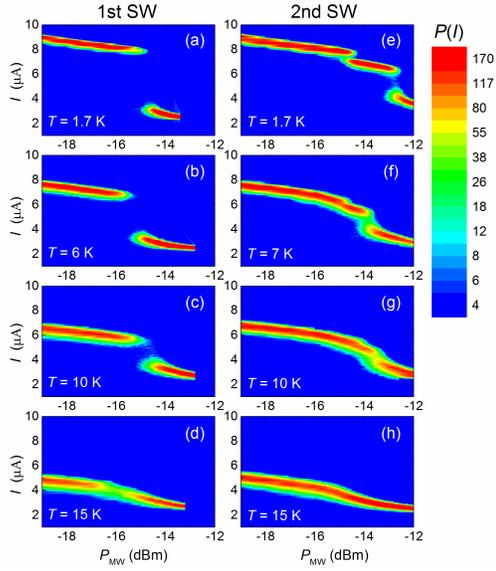}
\caption{
(Color online) (a)-(d) Density plots of the switching current versus the applied microwave power at $\omega/2\pi$=21.4~GHz for the 1st SW from 0.8~K to 15~K, (e)-(h) The same density plots for the 2nd SW. }
\label{f3}
\end{figure}

\begin{figure}[h]
\includegraphics[width=70mm]{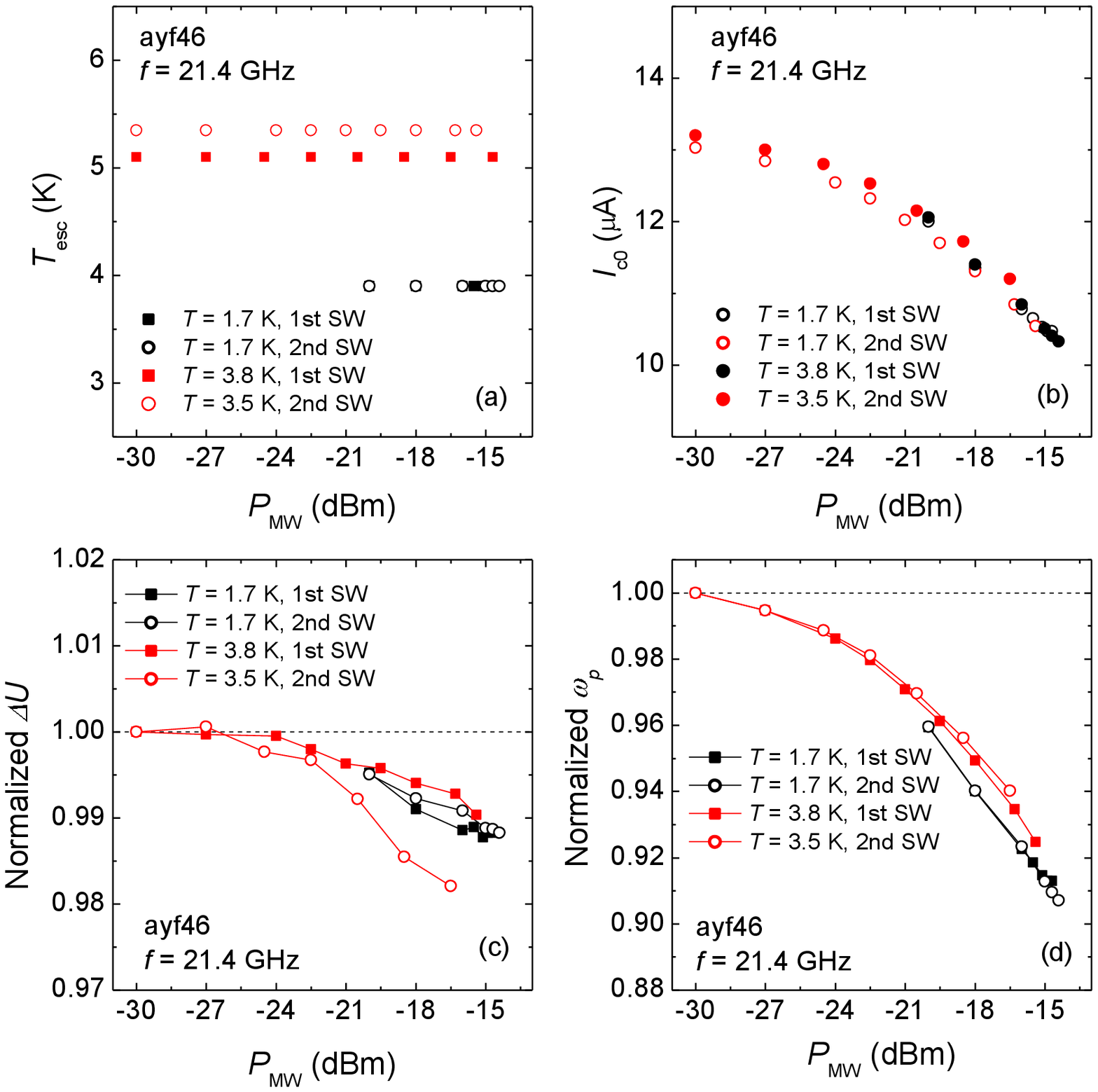}
\caption{
(Color online) (a) $T_{\rm esc}$, (b) $I_{\rm c0}$, (c) $\Delta U$ and (d) $\omega_p$ as a function of the applied microwave power. Here, $\Delta U$ and $\omega_p$ are given by a bias current at the peak of the switching current distribution. Solid and open symbols are the 1st SW and the 2nd SW, respectively. }
\label{f4}
\end{figure}


\begin{thebibliography}{9}
\bibitem{Inomata} K. Inomata, S. Sato, Koji Nakajima, A. Tanaka, Y. Takano, H. B.Wang, M. Nagao, H. Hatano, and S. Kawabata, Phys. Rev. Lett. {\bf 95}, 107005 (2005). 
\bibitem{Jin} X. Y. Jin, Y. Koval, A. Lukashenko, A. V. Ustinov, and P. M\"{u}ller, Phys. Rev. Lett. {\bf 96}, 177003 (2006).
\bibitem{Severev} S. Sevel'ev, A. L. Rakhmanov, and F. Nori, Phys. Rev. Lett. {\bf 98}, 077002 (2007); Erratum {\bf 98}, 269901 (2007). 
\bibitem{Fistul2} M. V. Fistul, Phys. Rev. B {\bf 75}, 014502 (2007).
\bibitem{Koyama} T. Koyama and M. Machida, Physica C {\bf 468}, 695 (2008). 
\bibitem{Kashiwaya} H. Kashiwaya, T. Matsumoto, H. Shibata, S. Kashiwaya, H. Eisaki, Y. Yoshida, S. Kawabata, and Y. Tanaka, J. Phys. Soc. Jpn. {\bf 77} 104708 (2008). 
\bibitem{Ota} K. Ota, K. Hamada, R. Takemura, M. Ohmaki, T. Machi, K. Tanabe, M. Suzuki, A. Maeda, and H. Kitano, Phys. Rev. B {\bf 79} 134505 (2009). 
\bibitem{Warburton} P. A. Warburton, S. Saleem, J. C. Fenton, M. Korsah, and C. R. M. Gronvenor, Phys. Rev. Lett. {\bf 103}, 217002 (2009). 
\bibitem{Nomura} Y. Nomura, T. Mizuno, H. Kambara, Y. Nakagawa, and I. Kakeya,  J. Phys. Soc. Jpn. {\bf 84} 013704 (2015).
\bibitem{Kakehi} D. Kakehi, Y. Takahashi, H. Yamaguchi, S. Koizumi, S. Ayukawa and H. Kitano, IEEE Trans. Appl. Supercond. {\bf 26}(3), 1800204 (2016). 
\bibitem{Kitano} H. Kitano, Y. Takahashi, D. Kakehi, H. Yamaguchi, S. Koizumi, and S. Ayukawa, J. Phys. Soc. Jpn. {\bf 85}, 054703 (2016).  
\bibitem{Takahashi} Y. Takahashi, D. Kakehi, S. Takekoshi, K. Ishikawa, S. Ayukawa, and H. Kitano, J. Phys. Soc. Jpn. {\bf 85} 073702 (2016).
\bibitem{Kitano2} H. Kitano, A. Yamaguchi, Y. Takahashi, D. Kakehi, and S. Ayukawa, J. Phys: Conference Series, {\bf 871} 012008 (2017).
\bibitem{Nomura2}  Y. Nomura, R. Okamoto, and I. Kakeya, Supercond. Sci Technol. {\bf 30} 105001 (2017).
\bibitem{Kitano3} H. Kitano, A. Yamaguchi, Y. Takahashi, S. Umegai, Y. Watabe, H. Ohnuma, K Hosaka, and D. Kakehi, J. Phys: Conference Series, {\bf 969} 012065 (2018).
\bibitem{Caldeira} A. O. Caldeira and A. J. Leggett, Phys. Rev. Lett. {\bf 46}, 211 (1981).
\bibitem{Fistul} M. V. Fistul, A. Wallraff, and A. V. Ustinov, Phys. Rev. B {\bf 68} 060504 (2003).
\bibitem{HFYu} H. F. Yu, X. B. Zhu, J. K. Ren, Z. H. Peng, D. J. Cui, H. Deng, W. H. Cao, Ye Tian, G.H. Chen, D. N. Zheng, X. N. Jing, Li Lu, and S. P. Zhao, New J. Phys. {\bf 15}, 095006 (2013).  
\bibitem{SXLi} Shao-Xiong Li, W. Qiu, S. Han, Y. F. Wei, X. B. Zhu, C. Z. Gu, S. P. Zhao, and H. B. Wang, Phys. Rev. Lett. {\bf 99}, 037002 (2007).
\bibitem{Kakizaki} Y. Kakizaki, J. Koyama, A. Yamaguchi, S. Umegai, S. Ayulawa, and H. Kitano, Jpn. J. Appl. Phys. {\bf 56} 043101 (2017).
\bibitem{Nomura3} Y. Nomura, T. Mizuno, H. Kambara, Y. Nakagawa, T. Watanabe, I. Kakeya, and M. Suzuki, J. Phys: Conference Series, {\bf 507} 012038 (2014).
\bibitem{Memo1} See Supplementary Material for details.
\bibitem{Suzuki} M. Suzuki, T. Hamatani, K. Anagawa, and T. Watanabe, Phys. Rev. B {\bf 85} 214529 (2012). 
\bibitem{Memo2} Note that we cannot determine the position of switched junctions in a stack of IJJs. However, we consider that the two junctions corresponding to the 1st and 2nd SWs are closely located, as discussed in Supplementary Material. 
\bibitem{Cleland} A. Cleland, J. Martinis, and J. Clarke, Phys. Rev. B {\bf 37} 5950 (1988). 
\bibitem{Grabert} H. Grabert, P. Olschowski, and U. Weiss, Phys. Rev. B {\bf 32} 3348 (1985); Phys. Rev. B {\bf 36} 1931 (1987). 
\bibitem{Martinis} J. Martinis, M. Devoret, and J. Clarke, Phys. Rev. Lett. {\bf 55}, 1543 (1985).
\bibitem{Kopietz} P. Kopietz and S. Chakravarty, Phys. rev. B {\bf 38}, 97 (1988). 
\bibitem{Takahashi_spl} The supplementary material in Ref. [\cite{Takahashi}].
\bibitem{HFYu2} H. F. Yu, X. B. Zhu, Z. H. Peng, W. H. Cao, D. J. Cui, Ye Tian, G.H. Chen, D. N. Zheng, X. N. Jing, Li Lu, and S. P. Zhao, S. Han, Phys. Rev. {\bf 81}, 144518 (2010).

\end{thebibliography}
\end{document}